# Number of New Top 2% Researchers from China and USA Over Time

Lei Liu[1], Song Yao[2], Kevin Liu[3]

[1]Division of Biostatistics, Washington University in St. Louis.

[2]Department of Marketing, Washington University in St. Louis

[3]Ladue Horton Watkins High School, St. Louis, MO

## Abstract

In this paper we compare the numbers of new top 2% researchers from China and USA annually since 1980. We find that the log ratio of the numbers decreases almost linearly over time. As early as 2009, the total number of new top 2% researchers across all subfields from China exceeds that of USA. In particular, such trend is more striking in many subfields, e.g., Engineering, Chemistry, and Enabling & Strategic Technologies.

## Section I. Introduction

In recent years the academic research capability increases dramatically in China (Xie et al. 2014). According the US National Science Foundation, China surpassed USA as the largest producer of scientific papers in 2017 (Tollefson 2018). Japan's National Institute of Science and Technology Policy (NISTEP), affiliated with its Ministry of Education, Culture, Sports, Science and Technology (MEXT), revealed that China has overtaken the United States for the first time in terms of the influence of research papers, in terms of the number of the top 10 percent of the most cited scientific papers in each research field (Nihon Keizai Shimbun (Nikkei) August 10,

[1] Corresponding to: Lei Liu. Email: lei.liu@wustl.edu

2021). In this article, we will compare China vs. USA on the number of new top researchers from 1980 to 2017 by the year of first publication (academic career start).

Our data come from the paper "Updated science-wide author databases of standardized citation indicators". In 2020 Ioannidis et al. provided two databases of top 2% researchers who received the most citations (i) in total (i.e., lifetime up to 2019) and (ii) in year 2019 alone. The databases include information on the top 2% researchers, including their name, subfield, year of first publication (career start), country, number of citations in total and in subfields. Such databases provide valuable information to compare the research capability of different countries over time.

We will use the database of researchers for the most citations in year 2019 alone. We choose this database over the other for two reasons: (i) it reflects the current research impact/significance of the publications; (ii) it is fairer to newer researchers since they have less total publications and citations. Using this database, we will compare the number of top 2% researchers from the currently top 2 countries in research capability: China vs. USA. We only count the researchers who started their career (published their first paper) since 1980 and up to 2017.

## Section II. Comparison In All Fields

All the analyses are conducted in Python. We first show the number of top 2% researchers from China and USA in Figure 1 from year 1980 to 2017. The X axis shows the year of top 2% researchers publishing their first paper (career start). We note that in 2009 the number of top researchers from China exceeded that from US, though the two curves are close more recently.

Naturally, the number of top 2% researchers decrease for newer researchers, which might result from 2 facts: (i) newer researchers have less publications by 2017; (ii) their papers are more recent so the impact has not been fully exerted in terms of citations. Therefore, the ratio of the number of top researchers may better reflect the research capability between the two countries.

In Figure 2 we show the ratio of the number of top researchers from USA vs. China over time across all subfields. It can been seen that the ratio keeps decreasing in an exponential way (note the Y axis is in log scale).

We run a linear model on the log ratio vs. year of first publication. We take calendar year 1980 as the baseline (year 0). Our model is

$$\log \text{ratio} = \beta_0 + \beta_1 \text{year}$$

The estimate of $\beta_0$ and $\beta_1$ is 3.46 (SE 0.11) and -0.13 (SE 0.0053). We can see that the decreasing trend is highly significant: the p-value for $\widehat{\beta_1}$ is 8.6×10$^{-24}$. Each calendar year increase is associated with 12% decrease in the ratio. We can transform it back to the original scale:

$$\text{ratio} = \exp(3.46 - 0.13 \text{ year}).$$

## Section III. Comparison in Subfields

We also calculate the ratio by subfields. There are a total of 20 subfields: 'Economics & Business', 'Physics & Astronomy', 'Information & Communication Technologies', 'Enabling & Strategic Technologies', 'Engineering', 'Clinical Medicine', 'Biology', 'Earth & Environmental Sciences', 'Biomedical Research', 'Psychology & Cognitive Sciences', 'Mathematics & Statistics', 'Agriculture, Fisheries & Forestry', 'Chemistry', 'Historical Studies', 'Social Sciences', 'Philosophy & Theology', 'Built Environment & Design', 'Visual & Performing Arts', 'Communication & Textual Studies', 'Public Health & Health Services'.

The plots are shown in Figure 3A to 3Q. It can be seen that most ratios decrease over time in a constant and steady pace. In a log scale, the decreasing trends are close to linear. The ratios become less than 1 in many subfields, especially in science and engineering. For example, as early as between 1995 to 2000, the number of top researchers from China already surpassed that from US in three subfields: Enabling & Strategic Technologies, Engineering, and Chemistry.

US still has advantage in the following subfields: Economics & Business, Clinical Medicine, Biomedical Research, Social Science, Public Health & Health Services. Further, the US dominates in Historical Studies, Philosophy & Theology, Psychology & Cognitive Sciences, Visual & Performing Arts, Communication & Textual Studies, though the numbers of top researchers are in general small in these subfields, so the ratios would be unstable over time and not shown in this paper.

## Section IV. Conclusion

In conclusion, although until recently China surpassed USA in terms of the total number of publications (in 2017) and high impact publications (in 2018), the number of young top researchers in China caught up with USA more than a decade ago (around 2009). In many scientific and engineering subfields China has a clear advantage. USA must devote more efforts and resources in education and research, in particular science and engineering, to enhance the quality and quantity of the new generation of researchers.

Finally, it is worth noting that we have not considered the translation of scientific publications to breakthrough innovation. In a report titled "why China is falling behind on breakthrough innovation" (Zha 2021), a leading Chinese scientist Yigong Shi raised the issue that "an appearance of prosperity, based merely on size and quantity of research publications, … would not necessarily lead to a boost in science and engineering". New policies have been stipulated to expedite such a translation process in China. Question remains on how to more accurately compare the scientific innovation between the two countries.

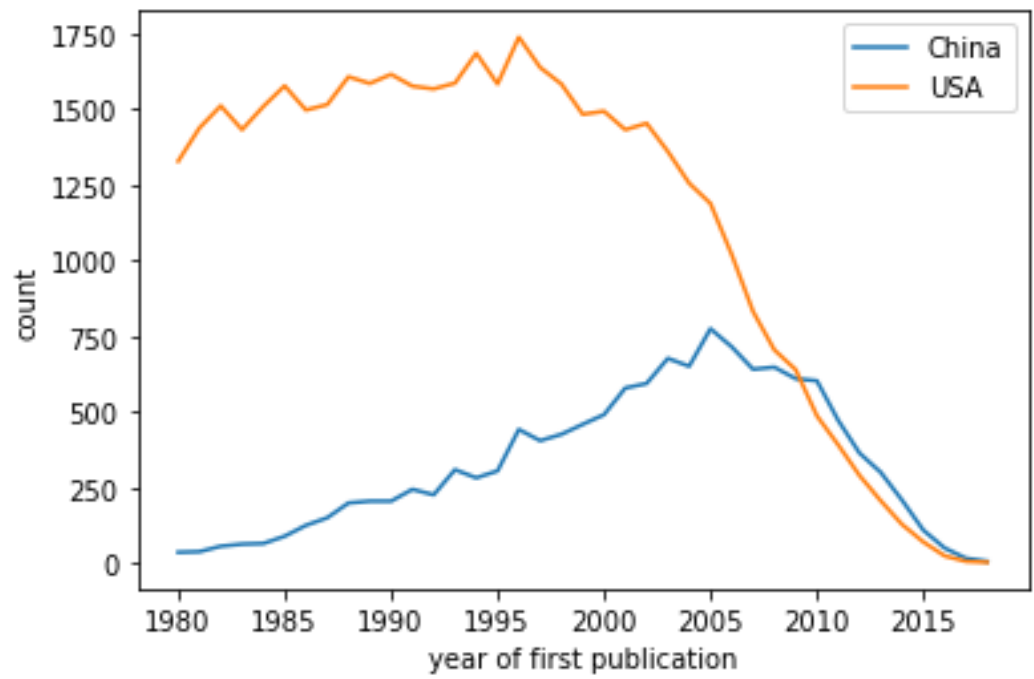

Figure 1. Number of top 2% reseachers from China and USA

Figure 2. Ratio of the number of top researchers from China vs. USA over time across all subfields

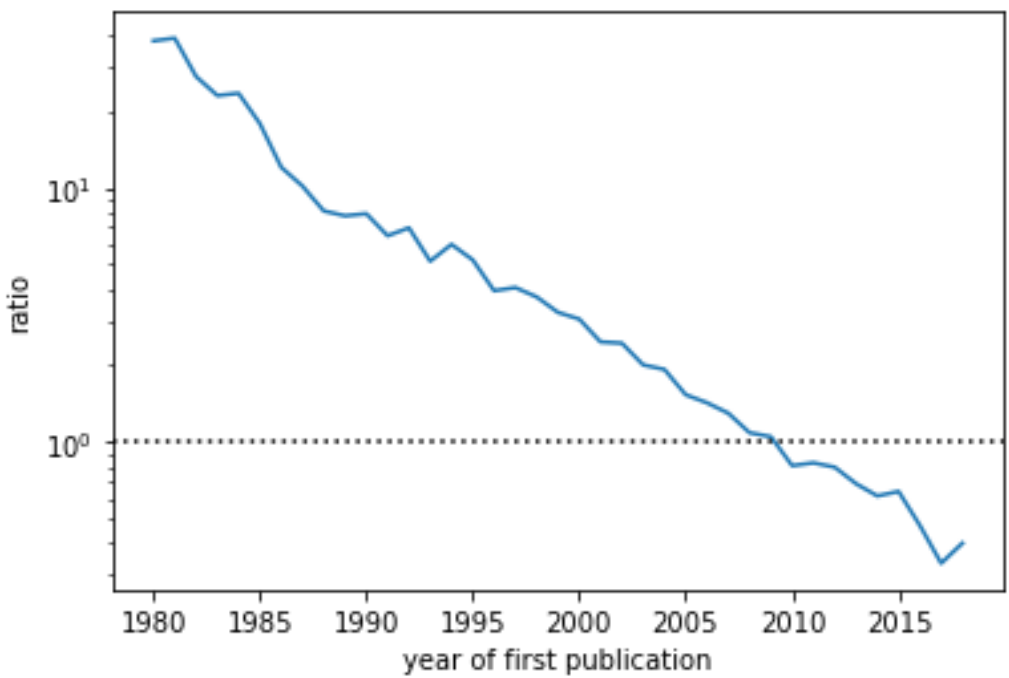

Figure 3A. Ratio of the number of top researchers from China vs. USA over time in subfield Economics and Business

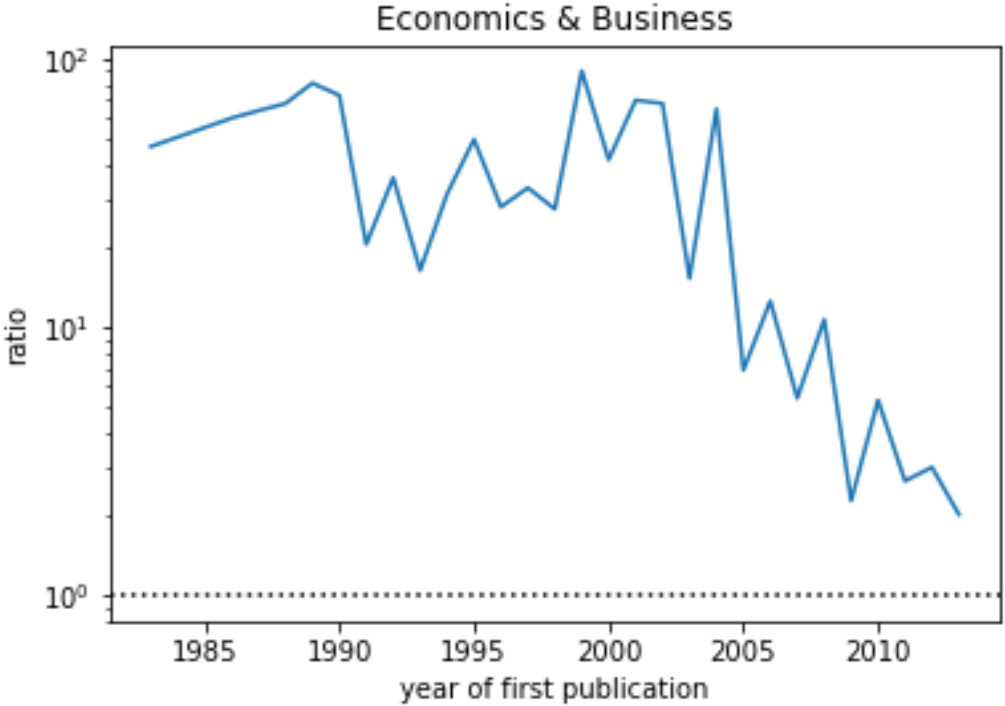

Figure 3B. Ratio of the number of top researchers from China vs. USA over time in subfield Physics and Astronomy

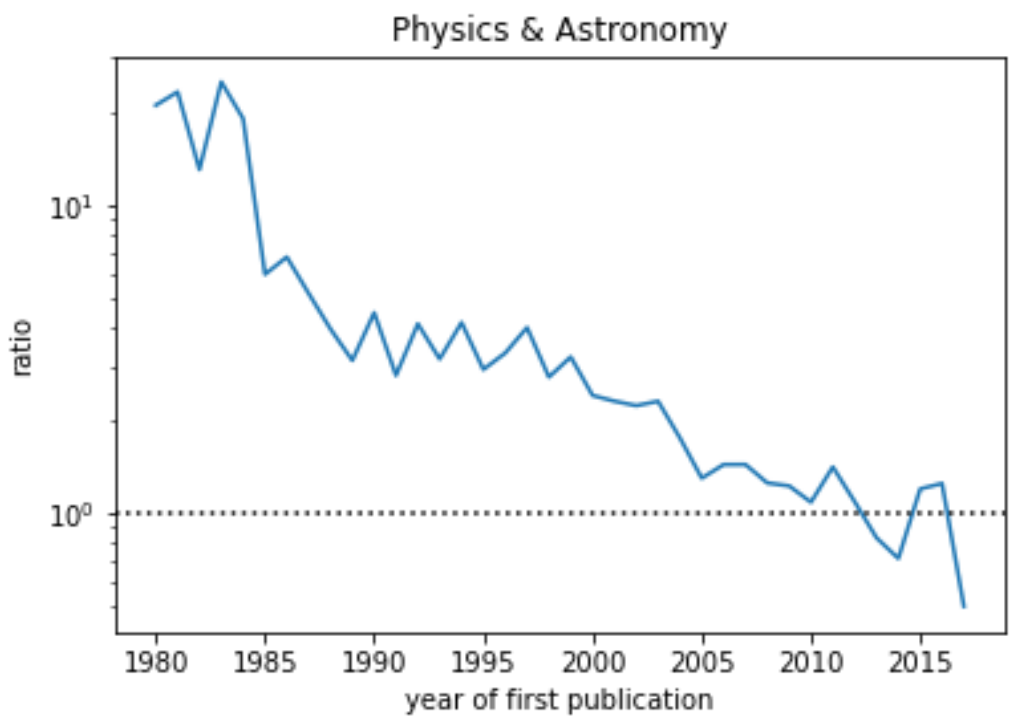

Figure 3C. Ratio of the number of top researchers from China vs. USA over time in subfield Information and Communication Technologies

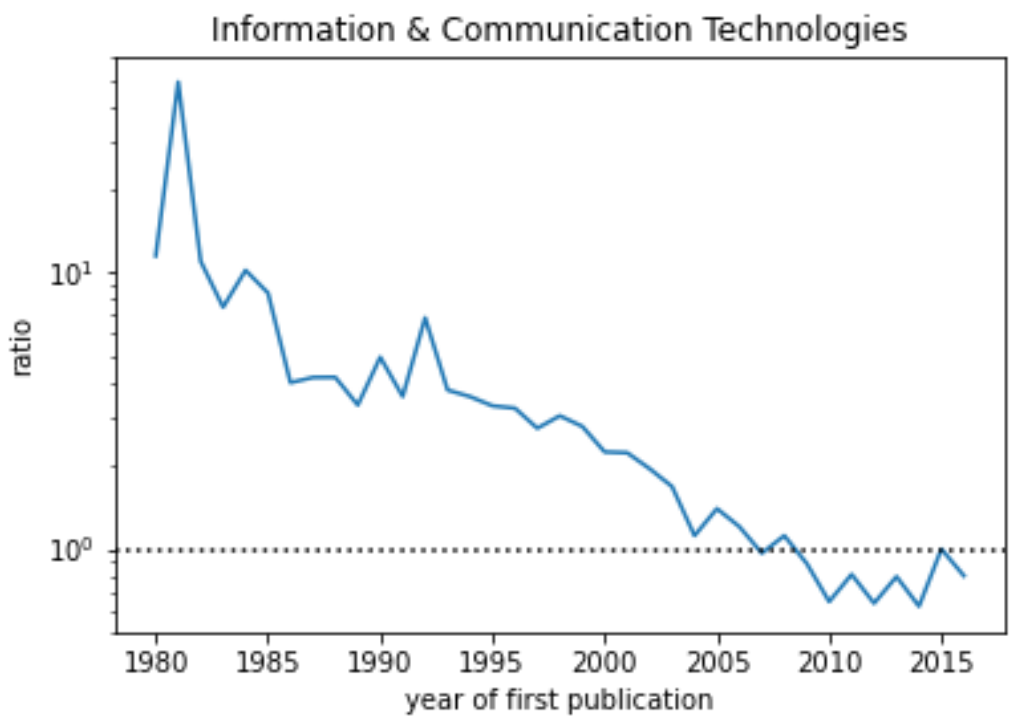

Figure 3D. Ratio of the number of top researchers from China vs. USA over time in subfield Enabling and Strategic Technologies

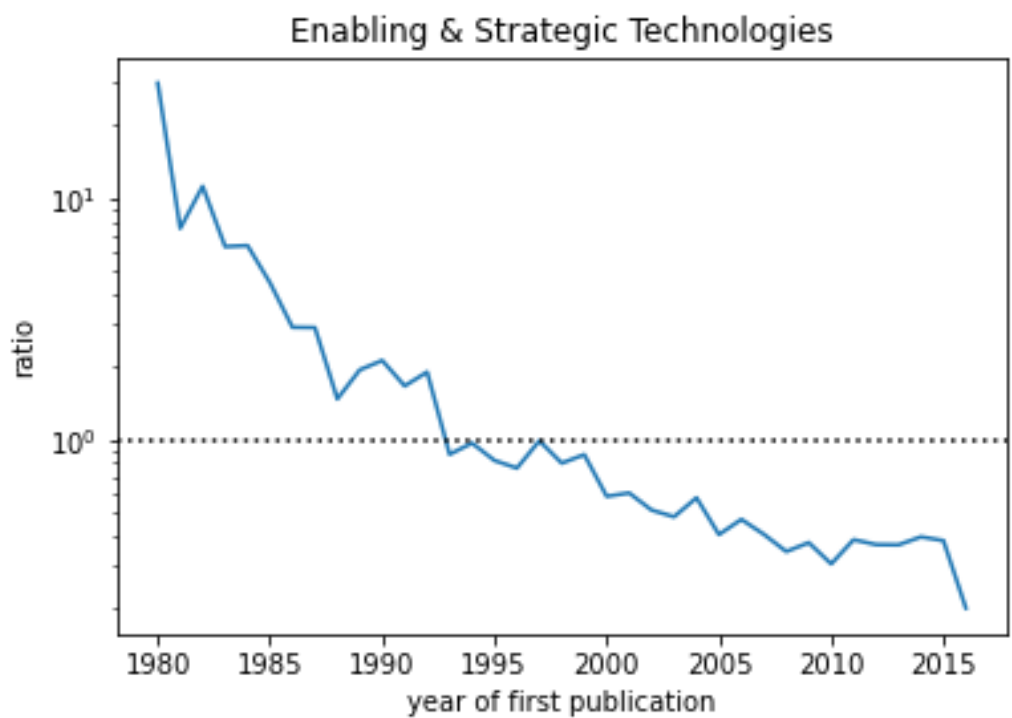

Figure 3E. Ratio of the number of top researchers from China vs. USA over time in subfield Engineering

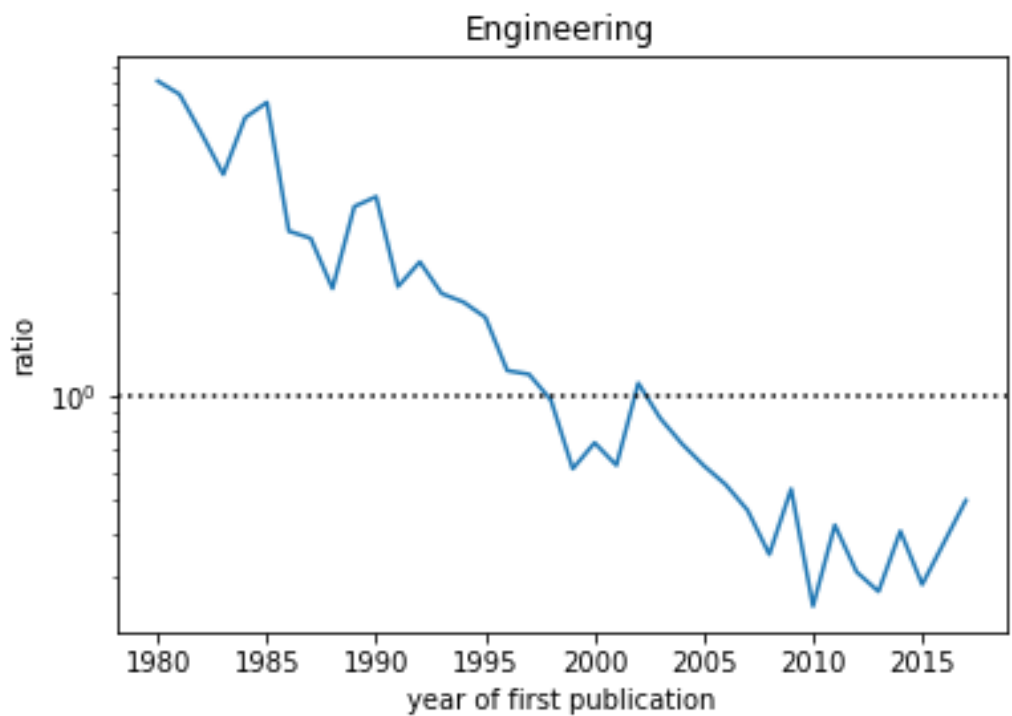

Figure 3F. Ratio of the number of top researchers from China vs. USA over time in subfield Clinical Medicine

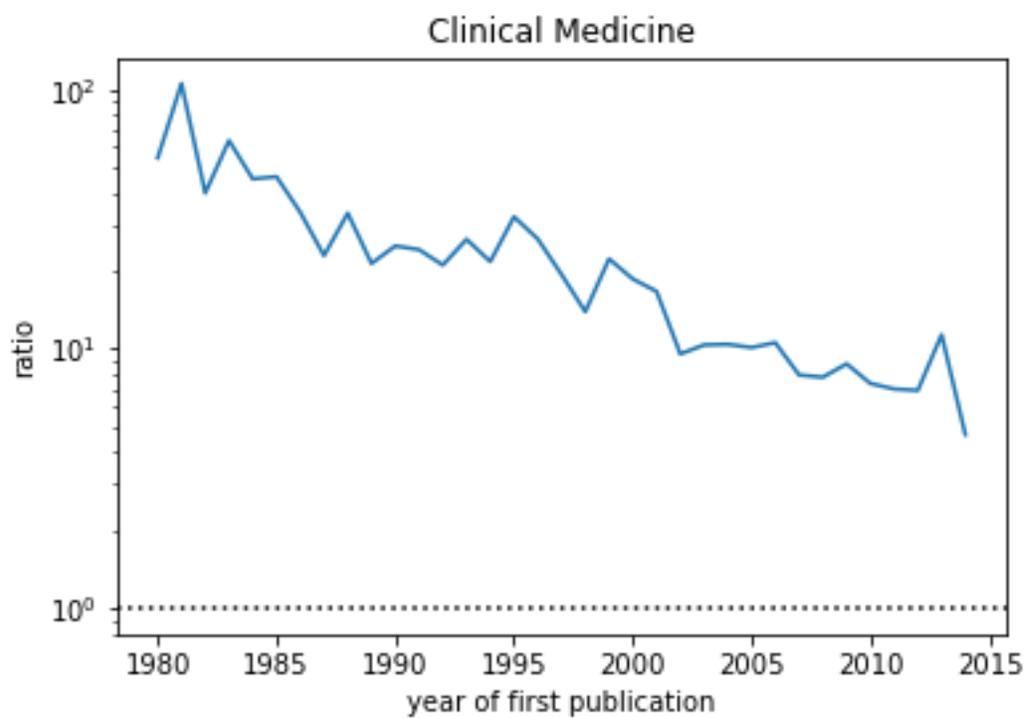

Figure 3G. Ratio of the number of top researchers from China vs. USA over time in subfield Biology

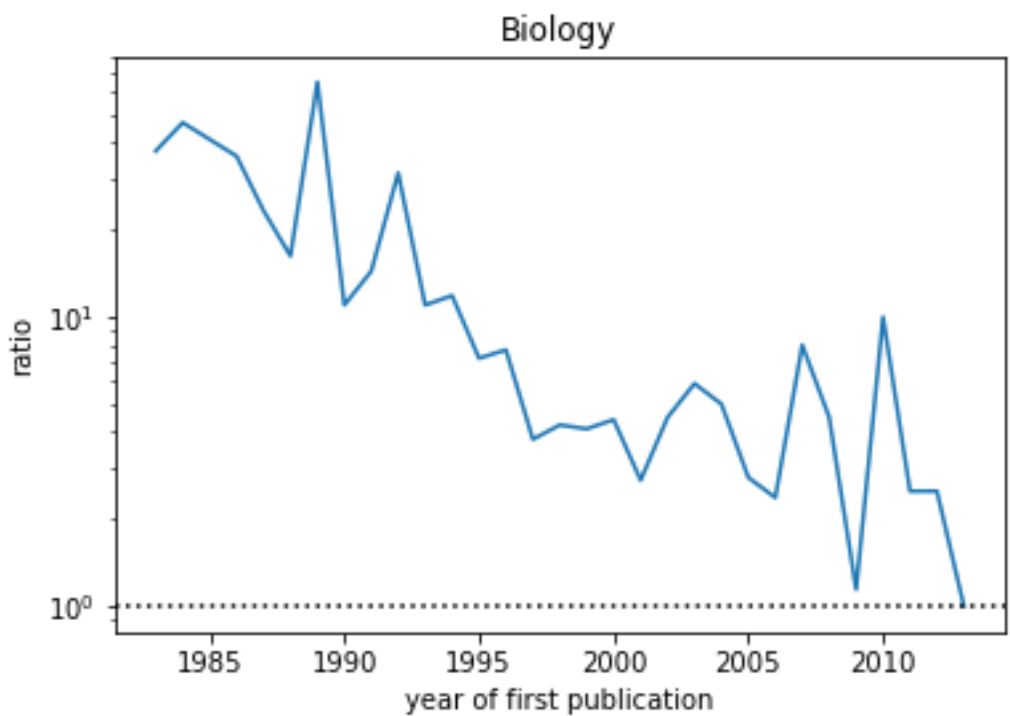

Figure 3H. Ratio of the number of top researchers from China vs. USA over time in subfield Earth and Environmental Sciences

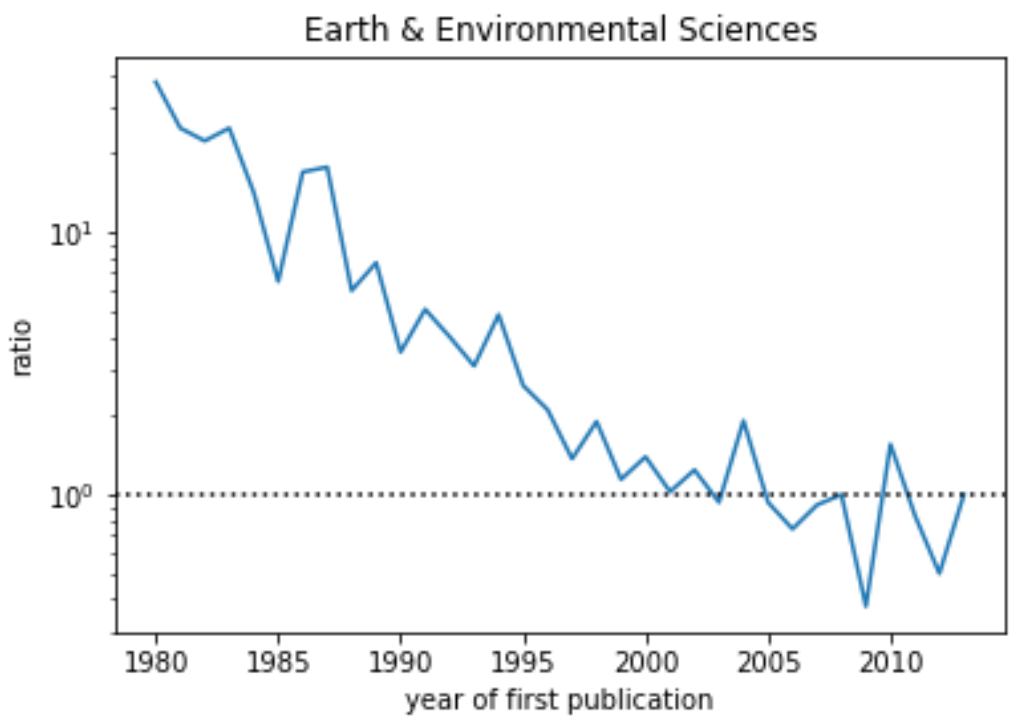

Figure 3I. Ratio of the number of top researchers from China vs. USA over time in subfield Biomedical Research

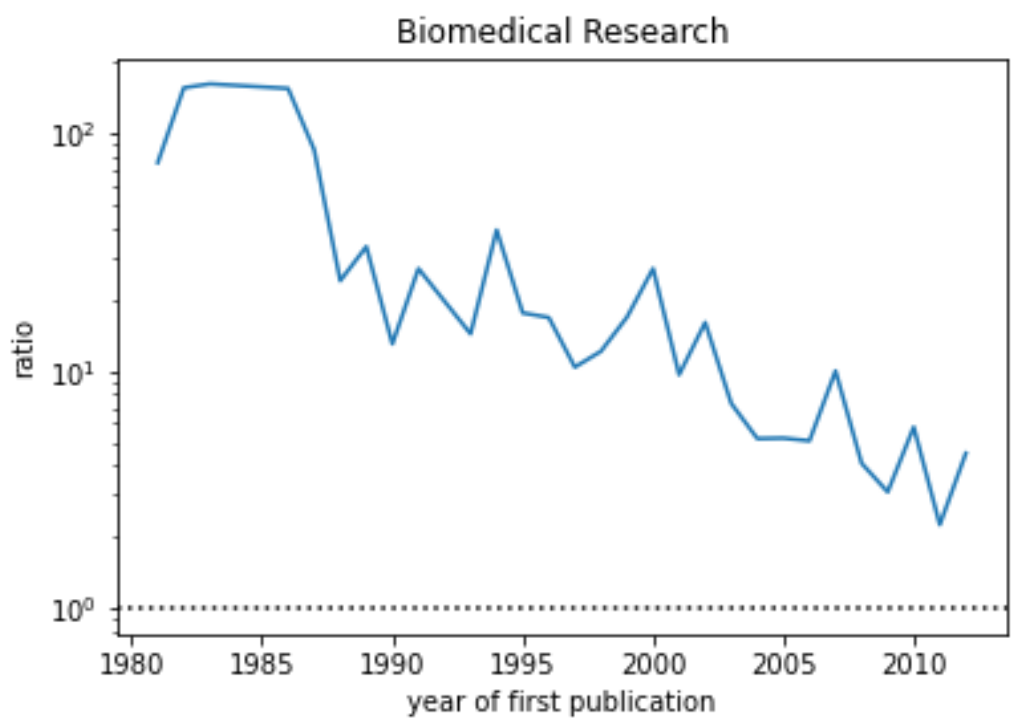

Figure 3J. Ratio of the number of top researchers from China vs. USA over time in subfield Psychology and Cognitive Sciences

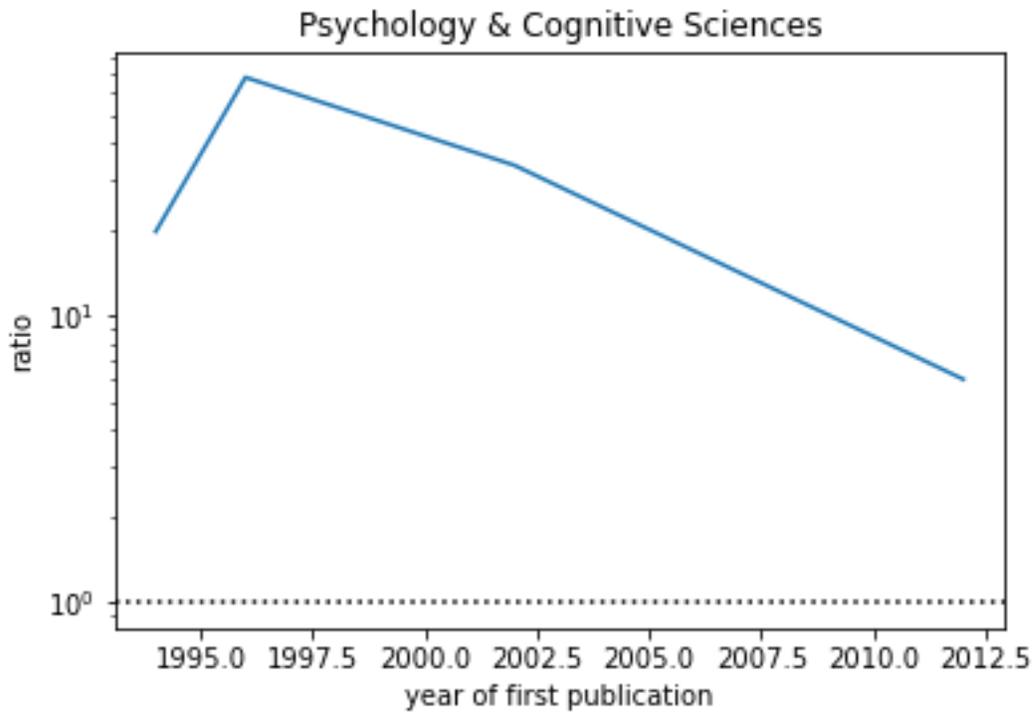

Figure 3K. Ratio of the number of top researchers from China vs. USA over time in subfield Mathematics and Statistics

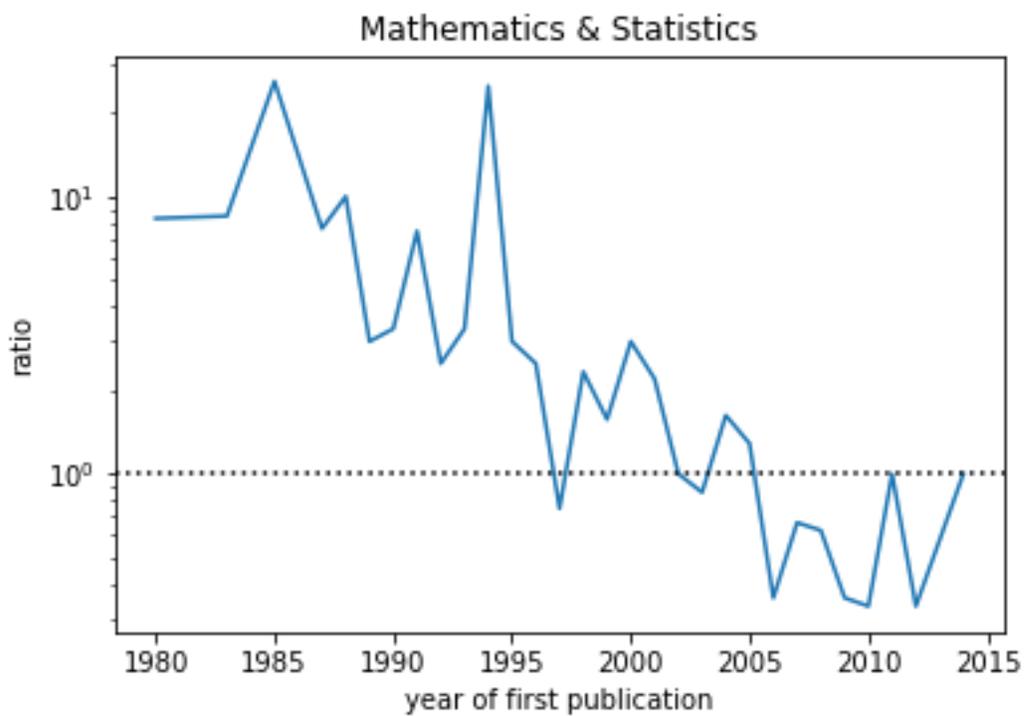

Figure 3L. Ratio of the number of top researchers from China vs. USA over time in subfield Agriculture, Fisheries, and Forestry

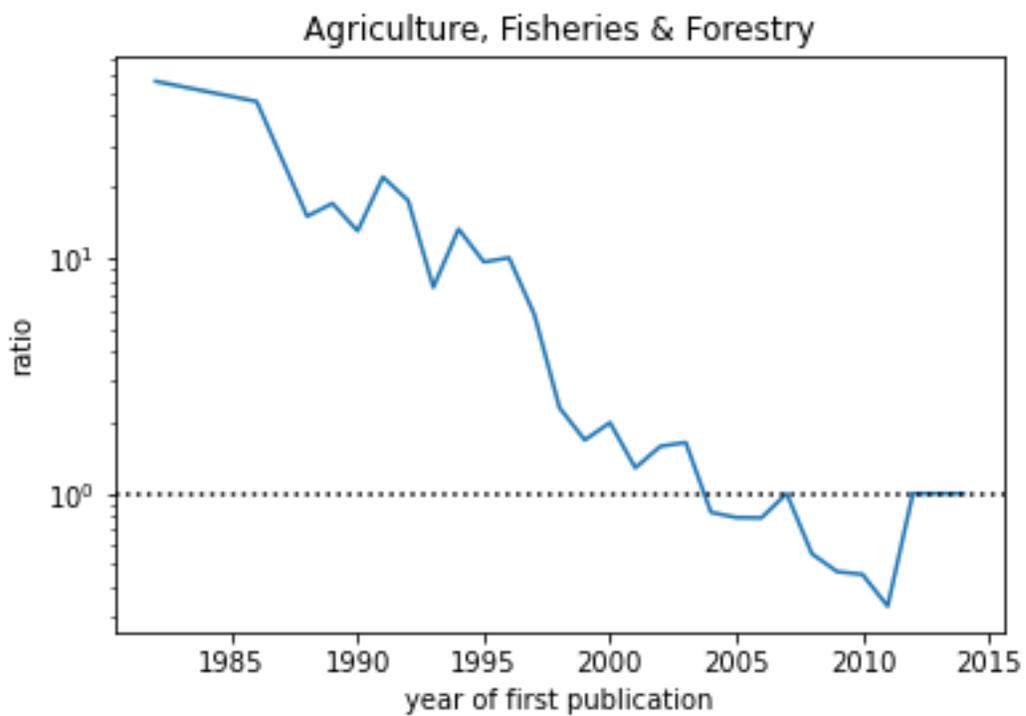

Figure 3M. Ratio of the number of top researchers from China vs. USA over time in subfield Chemistry

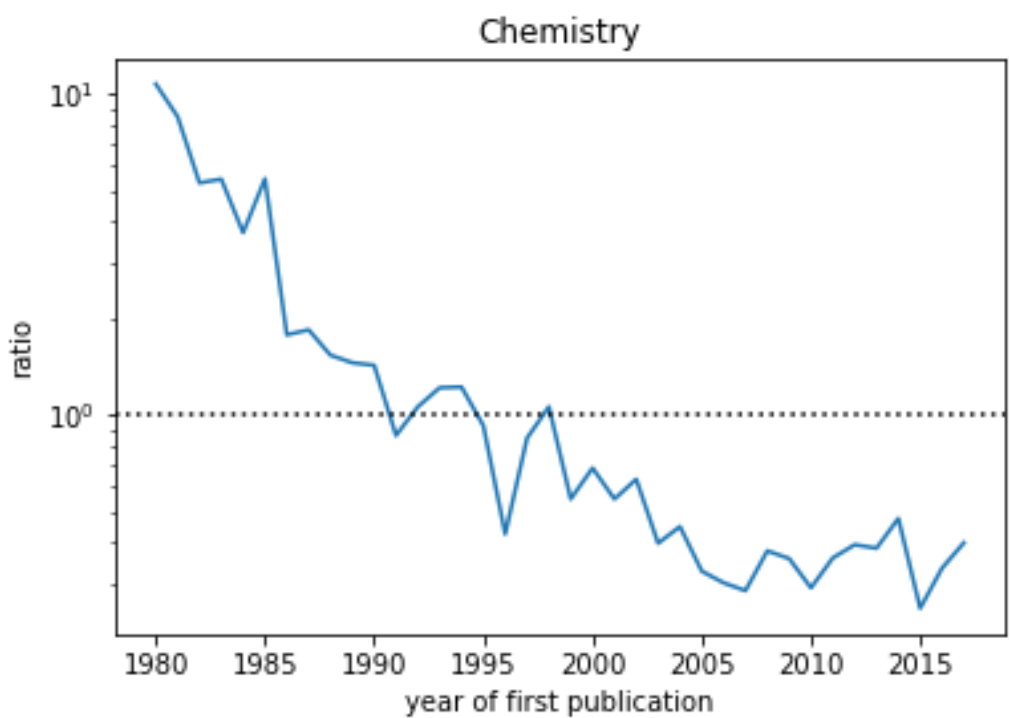

Figure 3N. Ratio of the number of top researchers from China vs. USA over time in subfield Social Sciences

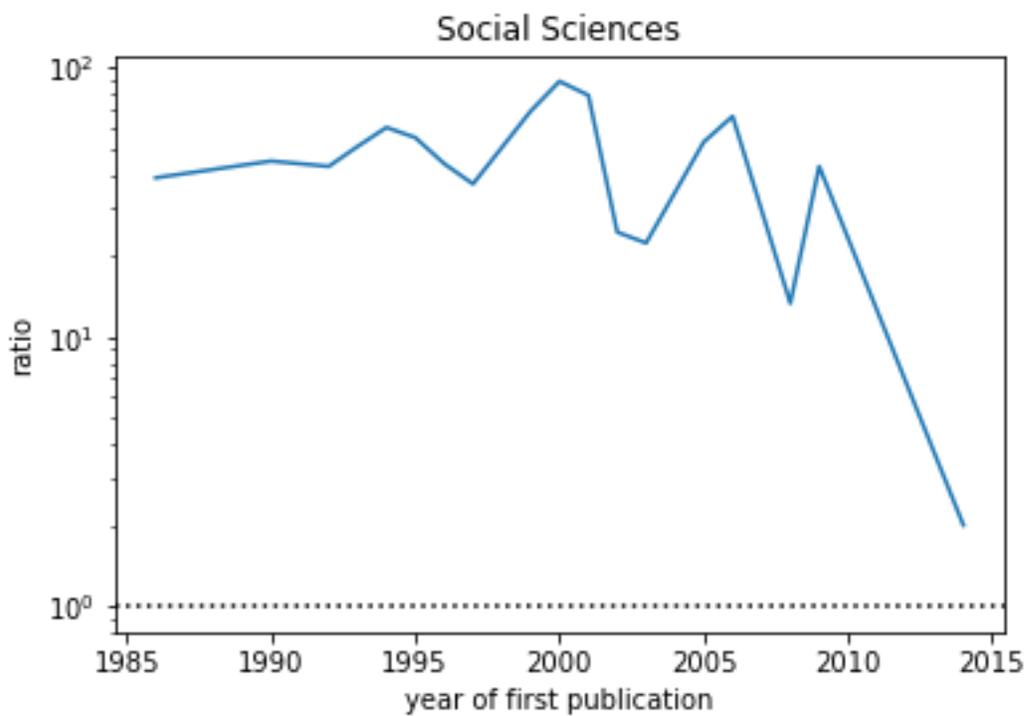

Figure 3O. Ratio of the number of top researchers from China vs. USA over time in subfield Built Environment and Design

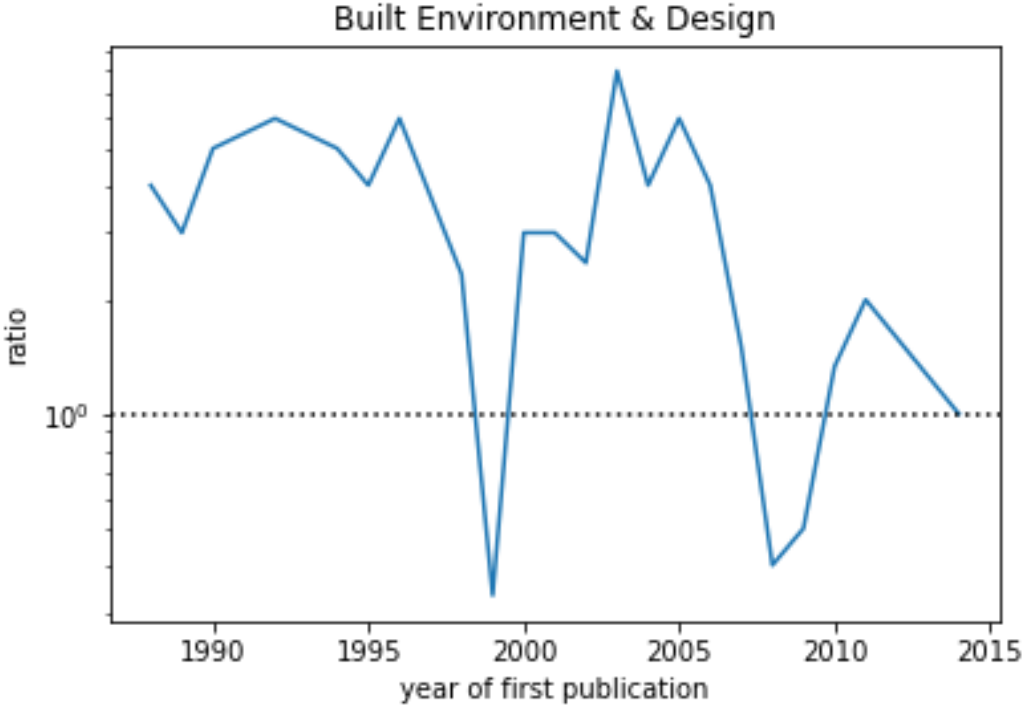

Figure 3P. Ratio of the number of top researchers from China vs. USA over time in subfield Communication and Textual Studies

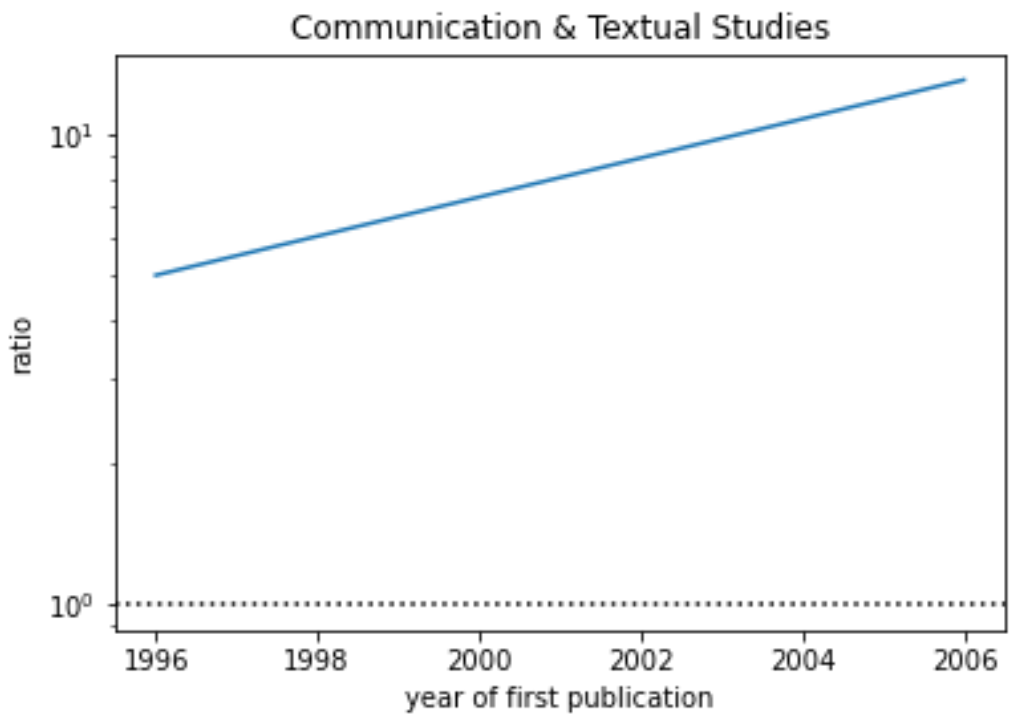

Figure 3Q. Ratio of the number of top researchers from China vs. USA over time in subfield Public Health and Health Services

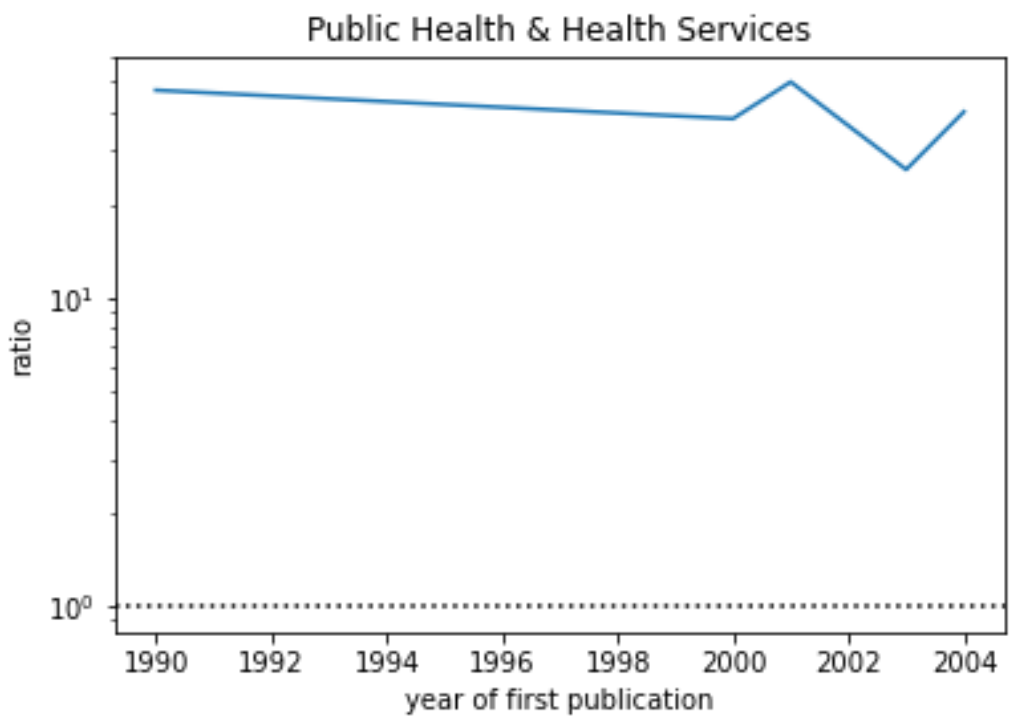